\documentclass[superscriptaddress,aps,nofootinbib,twocolumn]{revtex4}
\usepackage{graphicx}
\usepackage{amsmath,amssymb}

\def\bkR{{\rm I\kern-.17em R}}
\def\bkC{{\rm \kern.24em \vrule width.05em height1.4ex depth-.05ex \kern-.26em C}}
\def\to{\rightarrow}

\def\CQG{{\it Class. Quantum Gravity} }

\def\MPL{{\it Mod. Phys. Lett.} }

\def\PL{{\it Phys. Lett.} }
\def\PR{{\it Phys. Rev.} }
\def\PRL{{\it Phys. Rev. Lett.} }

\def\be{\beta}

\def\frac#1#2{{\textstyle{{#1}\over {#2}}}}

\def\lsim{\mathrel{\rlap{\lower4pt\hbox{\hskip1pt$\sim$}}
    \raise1pt\hbox{$<$}}}
\def\gsim{\mathrel{\rlap{\lower4pt\hbox{\hskip1pt$\sim$}}
    \raise1pt\hbox{$>$}}}
\def\sqr#1#2{{\vcenter{\vbox{\hrule height.#2pt
         \hbox{\vrule width.#2pt height#1pt \kern#1pt
         \vrule width.#2pt}
         \hrule height.#2pt}}}}

\def\laq{\raise 0.4 ex \hbox{$<$}\kern -0.8 em\lower 0.62 ex\hbox{$\sim$}}
\def\gaq{\raise 0.4 ex \hbox{$>$}\kern -0.7 em\lower 0.62 ex\hbox{$\sim$}}

\def\be{\begin{equation}}
\def\ee{\end{equation}}
\def\ba{\begin{eqnarray}}
\def\ea{\end{eqnarray}}

\def\dalemb#1#2{{\vbox{\hrule height.#2pt
        \hbox{\vrule width.#2pt height#1pt \kern#1pt \vrule width.#2pt}
        \hrule height.#2pt}}}

\def\dalemb#1#2{{\vbox{\hrule height.#2pt
        \hbox{\vrule width.#2pt height#1pt \kern#1pt \vrule width.#2pt}
        \hrule height.#2pt}}}

\def\gtorder{\mathrel{\raise.3ex\hbox{$>$}\mkern-14mu
             \lower0.6ex\hbox{$\sim$}}}
\def\ltorder{\mathrel{\raise.3ex\hbox{$<$}\mkern-14mu
             \lower0.6ex\hbox{$\sim$}}}

\begin{document}

\title{ Unparticle inspired corrections to the Gravitational Quantum Well}

\author{A. Alves\footnote{Email: andre.alves@ist.utl.pt}}

\vskip 0.3cm

\affiliation{Departamento de F\'\i sica, Instituto Superior T\'ecnico \\
Avenida Rovisco Pais 1, 1049-001 Lisboa, Portugal}

\author{O. Bertolami\footnote{Email: orfeu@cosmos.ist.utl.pt}}

\vskip 0.3cm

\affiliation{Departamento de F\'\i sica, Instituto Superior T\'ecnico \\
Avenida Rovisco Pais 1, 1049-001 Lisboa, Portugal}
\affiliation{Instituto de Plasmas e Fus\~ao Nuclear, Instituto Superior T\'ecnico \\
Avenida Rovisco Pais 1, 1049-001 Lisboa, Portugal}

\vskip 0.3cm

\date{July 2010}



\begin{abstract}


{ We consider unparticle inspired corrections of the type  ${ (\frac{R_{G}}{r})}^\beta$ to the Newtonian potential in the context of the gravitational quantum well. The new energy spectrum is computed and bounds on the parameters of these corrections are obtained from the knowledge of the energy eigenvalues of the gravitational quantum well as measured by the GRANIT experiment.}

\end{abstract}


\maketitle

\section{Introduction}

Unparticle physics \cite{georgi} is an interesting possible extension to the standard model (SM) beyond the TeV scale. Unparticle physics arises from the possibility of implementing scale invariance in the SM. This requires considering an additional set of fields,  the Banks-Zaks (BZ) fields, with a non-trivial IR fixed point. The interaction between SM and BZ fields is mediated by particles with a large mass, $ M_{*}$. This coupling can be written as
\be {\cal L}_{BZ} = {1\over M_*^k}O_{SM}O_{BZ}~~,
\label{BZ}
\ee 
where $O_{SM}$ denotes an operator with mass dimension $d_{SM}$ constructed from SM fields and 
$O_{BZ}$ is an operator with mass dimension $d_{BZ}$ constructed from the BZ fields.
At an energy scale, $\Lambda_{U}$,  the BZ operators mutate into unparticles operators ($O_{U}$), with non-integer scaling dimension $d_{U}$, such that:
\be \label{LagU}
{\cal L}_U={C_U \Lambda_U^{d_{BZ}-d_U}\over M_*^k}O_{SM}O_{U}~~,
\ee \noindent  
where $C_{U}$ is a coefficient function. This extension can involve the exchange of scalar,  vector,  tensor or even spinor unparticles between SM particles. Signatures for colliders \cite{cheng, cheng2} as well as other phenomenological aspects have been investigated \cite{greiner,lenz,liu,obNS,obPS,obP}. The exchange of unparticles can give rise to forces which lead to deviations from the inverse square law (ISL). Indeed, if $O_{U}$ is a rank-two tensor it couples to the stress-energy tensor $T_{\mu\nu}$ and it gives origin to a modification of Newtonian gravity leading to an effective potential,which in the non-relativistic limit and for $d_{U}\ne 1$, has the following form \cite{goldberg}:
\be 
V(r)=- G_{N} {M m \over r}  \left[1+\left ({R^{T}_{G} \over r}\right)^{2 d_{U}-2}\right]~,
\label{VUT}
\ee 
where $G_{N}$ is the Newton's gravitational constant and $R^{T}_{G}$ is the characteristic length for which this ``ungravity" interactions become significant and is defined to be

\ba 
\label{R_G} 
R^T_G & = &{1\over \pi \Lambda_U} \left({M_{Pl} \over M_*}\right)^{1/(d_U-1)} \times \\ \nonumber && \left[ {2 (2-\alpha) \over \pi} {\Gamma(d_U+{1 \over 2})\Gamma(d_U-{1 \over 2}) \over \Gamma(2d_U)}\right]^{1/
(2d_U-2)}~~,
\ea 
where $M_{Pl}=1.22 \times 10^{19} GeV$ is the Planck mass and $\alpha$ is a constant dependent on the type of the propagator in question ($\alpha=1$ for the graviton case). The case $d_{u} < 1$ leads to forces which fall slower than gravity, which for large distances 
can be directly tested experimentally. Actually, torsion-balance experiments that search for power-law modifications of the inverse 
square law (ISL) have been performed and down to distances of around 0.05 mm no significant deviations have been found 
\cite{adelberger}. More specifically, for integers values of $\beta \equiv 2 d_{u}-2$, i.e. $\beta =1, 2, 3, 4$, positive corrections to the ISL are constrained to be smaller than $4.5 \times 10^{-4}$ than Newtonian gravity for $\beta=1$ and smaller than $1.5 \times 10^{-5}$ 
for $\beta=4$ \cite{adelberger}. As described above, unparticles exchange yields contributions to the ISL that correspond to non-integers of $\beta$ and, as will be seen, the GRANIT experimemtal results is particularly sensitive to values of $\beta \ltorder 0$, 
for a specific set of ranges (see below). 

If we consider a vector unparticle exchange \cite{Des}, the potential for a coupling between a vector unparticle and a baryonic (or leptonic) current $J_{\mu}$ combined with the gravitational potential is given by \cite{Des}:
\be 
V(r)=- G_{N} {M m \over r}  \left[1-\left ({{R^{V}_{G}} \over r}\right)^{2 d_{u}-2}\right]~,
\label{VUV}
\ee 
where ${R^{V}_{G}}$ is slightly different from ${R^{T}_{G}}$, but is clearly the characteristic length scale of this vector exchange. Notice that the vector contribution leads to a subtractive contribution to the potential given the repulsive nature of the vector unparticle exchange.

In what follows we consider the linear approximation of potentials, Eqs. (\ref{VUT}) and (\ref{VUV}), as corrections to the Newtonian potential.  In that we assume unparticle inspired corrections acting on scales away from the range the unparticle modifications were originally envisaged, as discussed above. In order to maximize the effects of the unparticle inspired modifications we assume that $R^{T,V}_{G} \gtrsim R_{E}$, where $R_{E}$ is Earth's radius. These assumptions allow to turn the corrections to the Newtonian potential into corrections on the gravitational quantum well (GQW) spectrum as measured by the GRANIT experiment \cite{granitN}. As will be seen, available data allow for obtaining bounds for parameters $R^{T}_{G}$, $R^{V}_{G}$ and $\beta = 2 d_{u}-2$.

\section{The Gravitational Quantum Well}

Let us now briefly discuss the GQW \cite{Griffths} and the data obtained by the GRANIT experiment \cite{granitN}. 
The experimental realization of the GQW  involves ultra-cold neutrons under the action of Earth's gravitational field, $\overrightarrow{g} =-g \overrightarrow{e_{z}}$ and an horizontal ``mirror" placed at $z=0$. This 2-dimensional problem can be reduced to an 1-dimensional quantum mechanical problem in the z-direction given that the neutrons' motion in the horizontal direction is free. The vertical motion of a bouncing particle at the quantum level is described by Schr\"odinger eigenvalue problem $\hat{H}\psi_{n}(z)=E_{n}\psi_{n}(z)$ with the 
Hamiltonian operator
\be 
\hat{H}= - {\hbar^2 \over 2 m}{d^2 \over dz^2} + mgz ~.
\label{gqw1} 
\ee
The solution to the problem is given in terms of regular Airy functions \cite{flugge}
\be
\psi_{n}(z)= N_{n}Ai [\theta z + \alpha_{n}] ~, 
\label{gqw2}
\ee
with energy eigenvalues
\be
E_{n}=- \left({mg \over \theta}\right) \alpha_{n}~,
\label{gqw2.1} 
\ee
where the normalization factor, $N_{n}$ and $\theta$ are given by
\be 
N_{n}= \left[ \int_{0}^{+\infty} (Ai [\theta z + \alpha_{n}])^2 dz \right]^{-\frac{1}{2}}~,
\label{gqw2.5} 
\ee

\be 
\theta= \left({2gm^2  \over \hbar^2}\right)^{1/3}~,
\label{gqw3} 
\ee
$\alpha_{n}$ being the $n$th zero of the regular Airy function Ai(z). Thus, 
neutrons bounce at the classical turning points corresponding to heights
\be 
h_{n}={E_{n} \over mg}=-{ \alpha_{n} \over \theta}~.
\label{gqw4} 
\ee
The first two zeros of Ai(z), $\alpha_{1}=-2.338$ and $\alpha_{2}=-4.088$, determine the first two critical heights, and through Eq. (\ref{gqw4}), the corresponding energy eigenvalues: 
\ba\label{gqwLim}
h^{th}_{1}= 13.7 (\mu m),~~ E_1=1.407 (peV),\\
h^{th}_{2}= 24.0  (\mu m),~~ E_2=2.461(peV). \nonumber
\ea

In the GRANIT experiment, ultra-cold neutrons with a mean horizontal velocity of $\langle v\rangle \simeq 6.5~ms ^{-1 }$  \cite{granitN}  freely move in the Earth's gravitational field. This setup mimics the GQW, meaning that the energy spectrum of 
neutrons under the action of gravity is quantized in the direction of the gravitational field and the probability of observing particles at a given height will be maximum at the classical turning points $h_{n}={E_{n}/ m g}$. The minimum energy is found to be $1.40 \times 10^{-12}$ eV, corresponding to a vertical velocity of $1.7~ cm s^{-1 }$ \cite{granitN}. The limit in accuracy is provided by the uncertainty principle and corresponds to an energy resolution of $10^{-18}$ eV, which could be achieved if the neutrons were  confined their whole lifetime, $\tau \simeq (885.7 \pm 0.8)~ s$ \cite{pdg2008}, within the experimental apparatus. 

Besides the ground states, three excited states were determined, although with reduced accuracy. The first two  measured position levels are:
\ba\label{granitLim}
h^{exp}_{1}= 12.2 \pm 1.8(syst) \pm 0.7 (stat) (\mu m), \\
h^{exp}_{2}= 21.6 \pm 2.2(syst) \pm 0.7 (stat) (\mu m). \nonumber
\ea

The GRANIT setup offers the opportunity to confront  observations with various theoretical models such as noncomutative geometry \cite{obNoncommu}, existence of an intrinsic minimal length \cite{brau}, the presence of extra dimensions contributions \cite{buisseret} and the effect of Yukawa type interactions \cite{obWEP}. Notice that the contributions of extra dimensions to the Newtonian potential is of the form Eq. (\ref{VUT}), however only for integer values of $\beta=2d_{U} -2$.

\section{The Modified Gravitational Quantum Well}

We analyze now the energy spectrum of the unparticle inspired potential 
\ba\label{modpot}
V(r)=- G_{N} {M m \over r}  \left[1\pm \left ({R_{G} \over r}\right)^{\beta}\right] ~~,
\ea
where we assume that $R^{T}_{G}=R^{V}_{G}=R_{G}$. We expand this potential around $R_{E}$ for $r=R_{E}+z$, keeping only the linear term in $z$. Thus, with respect to the potential $V=mgz$, we get that the unparticle inspired modification amounts to the change:
\be \label{modpotg}
g \to g '=g  \left[1 \pm (\beta+1) \left({R_{G} \over R_{E}}\right)^{\beta}\right] ,
\ee
where the plus sign denotes tensor unparticle exchange, while the minus sign vector unparticle exchange. Therefore, we can write for the energy eigenvalues of the modified GQW as
\ba 
E'_{n}=- \left({mg' \over \theta}\right) \alpha_{n}
\label{enmod} 
\ea
Through Eq. (\ref{enmod}) we can constrain $\beta$  for fixed  values of $R_{G}$ using the available GRANIT data.

\section{Results}

Next we analyze the effect of the unparticle corrections to the GQW energy spectrum for different values of $R_{G}$ according to the choice exhibited in Table 1. 

\begin{figure}[]
\includegraphics[width=8cm]{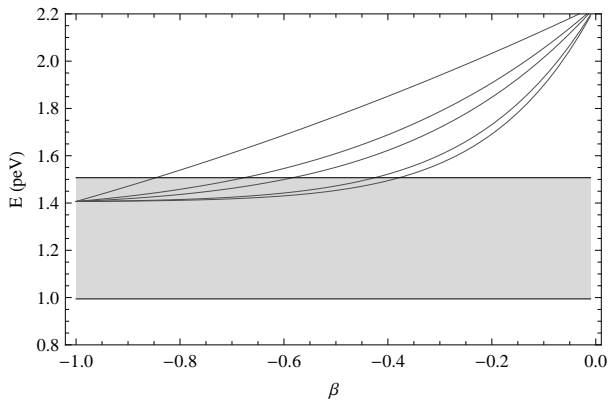} 
\caption[fig:resultsgto1]{First energy eigenvalue $E'_1$ for $R_{G} / R_{E}=1.5, 5, 10, 50, 100$ (from top to bottom) as a function of  $R_{G}$ and 
$\beta$ as modified 
by tensor unparticle exchange interactions. The grey region corresponds to the experimental results $E_{1} \pm \Delta E_{1}$.} 
\label{fig:resultsgto1}
\end{figure}

\begin{figure} [] 
\includegraphics[width=8cm]{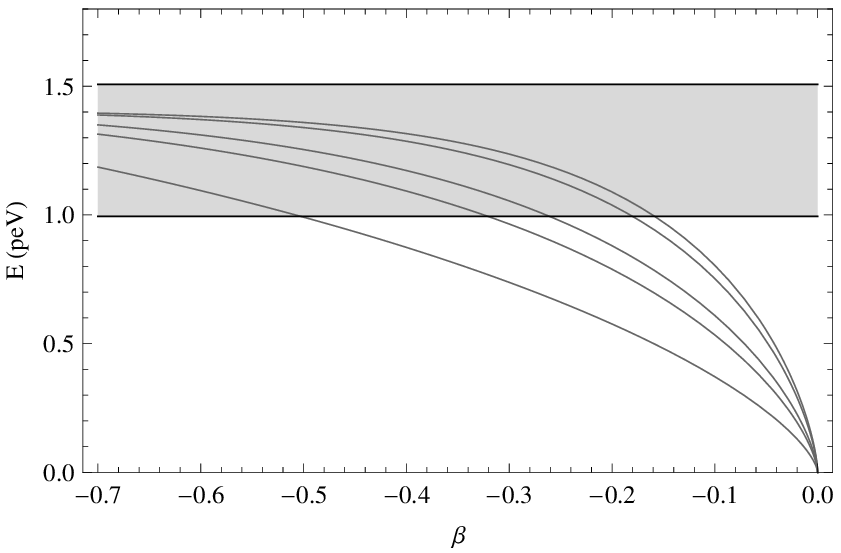}
\caption[fig:resultsgvo1]{First energy eigenvalue $E'_1$ for $R_{G}/R_{E}=100, 50, 10, 5, 1.5$ (from top to bottom) as a function of  $R_{G}$ and 
$\beta$ as modified 
by vector unparticle exchange interactions. The grey region corresponds to the experimental results 
$E_{1} \pm \Delta E_{1}$.}
\label{fig:resultsgvo1}
\end{figure}

For $R_{G} >  R_{E}$, the source of the new interaction is the whole Earth. Thus, using Eq. (\ref{enmod}), and GRANIT data we depict in Figs. \ref{fig:resultsgto1} and  \ref{fig:resultsgvo1} the contributions of tensor and vector unparticle exchange, respectively.
We test five different values of $R_{G}$, and vary continuously $\beta$ within the range that matches the  experimental measurement of the first energy level $E_{1} \pm \Delta E_{1}$. The intersection with the experimental error bars determines the lower and upper bounds for $\beta$. The results are summarized in Table 1.\\

For $R_{G} < R_{E}$, form factors must be introduced to account for the fact that only a shell of radius $R_{G} $ around the experiment is relevant . From computations that can be found in Ref. \cite{buisseret}, the resulting corrections to the ISL are too small and yield no significant bounds on the parameters $R_{G}$ and $\beta$.


\begin{table}[]
{
\renewcommand{\tabcolsep}{0.5cm}

\caption{Table of values for $\beta$ as a function of $R_{G}$ for the first energy level as modified by Eq. (\ref{enmod}). Values of $\beta$ greater than the ones indicated are excluded.}

\begin{tabular}[t] {c | c | c  }

$(R_{G} / R_{E}) $ &  \multicolumn{2}{c} {$\beta$} \\ 
\hline \hline
  &$ Vector $&$ Tensor $\\
\hline 
100  &   -0.158487 & -0.378236 \\
50  &  -0.179961& -0.425184  \\
10 & -0.260721 &  -0.582992\\
5  &  -0.320562 &  -0.676435\\
1.5 &  -0.50271 &  -0.846465  \\
\hline
\end{tabular}}\\ 
\end{table}

\vspace{0.5cm}

\section{Conclusions}

In this work we have considered unparticle inspired modifications to the ISL of the form of Eq. (\ref{modpot}). These corrections affect the GQW spectrum according to Eqs. (\ref{modpotg}) and (\ref{enmod}) and, as discussed in the text, the GRANIT experiment is particularly 
sensity to $\beta \ltorder 0$, which allows to scrutinize a region in the parameter space that is complementary to one tested by 
torsion-balance experiments. 
We find that depending on  the parameters $R_{G}$ and $\beta$, corrections for $ E'_{1}$ when compared with $E_{1}$ can be around 40\% for vector exchange and 10\% for tensor exchange. This asymmetry is due to the fact that the GRANIT experimental 
data for the upper and lower values for the critical heights, $ h^{exp}$ is not symmetric for $R_{G} >  R_{E}$.
Our results show that certain combinations of values of $R_G$ and $\beta$ are already excluded by the data and clearly an improvement on the GRANIT experiment may allow to exclude entirely certain ranges of unparticle-like contributions to the GQW. Given that the improvement of the GRANIT spectrometer is on the way (see e.g. http://lpsc.in2p3.fr/Indico/conferenceDisplay.py\break?confId=371), prospects to further test unparticles-type corrections to the GQW, as analyzed here, are rather realistic.

\subsection*{Acknowledgments}

The authors would like to thank  Valery Nesvizhevsky and Konstantin Protosov for discussions on the GRANIT experiment.

\vspace{0.3cm}

\noindent 


\end{document}